\documentclass[a4,11pt]{article}

\usepackage{amsfonts}
\usepackage[colorlinks=true]{hyperref}

\usepackage{amssymb}
\usepackage{amsmath}
\usepackage{graphicx}
\usepackage{relsize}

\usepackage{url}

\usepackage[linesnumbered,vlined,algoruled,noresetcount]{algorithm2e}
  \SetVlineSkip{0.0ex}\DontPrintSemicolon
  \SetKw{Halt}{halt}
  \SetKw{Break}{break}

\newcommand{\strong}{{\sf strong}}

\newcommand{\weak}{{\sf weak}}

\newcommand*{\ow}{\textsc{1}}
\newcommand*{\owdfa}{\ow\textsc{dfa}}

\newcommand*{\la}[1]{${#1}$-limited automaton}  
\newcommand*{\las}[1]{${#1}$-limited automata}  
\newcommand*{\onel}{\la{1}}
\newcommand*{\twol}{\la{2}}

\newcommand{\leftend}{{\mathord{\vartriangleright}}}
\newcommand{\rightend}{{\mathord{\vartriangleleft}}}

\newcommand{\mystackrel}[2]{%
  \mathrel{\vbox{\offinterlineskip\ialign{%
    \hfil##\hfil\cr
    $\scriptstyle#1$\cr
    $#2$\cr
}}}}

\newcommand{\mystackreg}[2]{%
  \mathrel{\vbox{\offinterlineskip\ialign{%
    \hfil##\hfil\cr
    $\scriptstyle#1$\cr
    \noalign{\kern1pt}
    $#2$\cr
}}}}

\newcommand{\moveL}{\mathrel{\scalebox{1}[0.9]{$\dashleftarrow$}}}
\newcommand{\moveR}{\mathrel{\scalebox{1}[0.9]{$\dashrightarrow$}}}
\newcommand{\moveWL}[1]{\mathrel{\mystackreg{\,#1}{\longleftarrow}}} 
\newcommand{\turnL}[2]{\mathrel{_{#2}\!\!\mystackrel{~\,#1}{\scalebox{1.6}[0.9]{$\hookleftarrow$}}}} 
\newcommand{\turnR}[2]{\mathrel{\mystackrel{#1~}{\scalebox{1.6}[0.9]{$\hookrightarrow$}}\!\!_{\!#2}}} 
\newcommand{\checkL}[1]{\mathrel{_{#1}\!\!\reflectbox{\scalebox{1.6}[0.9]{$\mapsto$}}}}

\newcommand{\openA}{\makebox[1.4ex][c]{\footnotesize\sf(}}
\newcommand{\closedA}{\makebox[1.4ex][c]{\footnotesize\sf)}}
\newcommand{\openB}{\makebox[1.4ex][c]{\footnotesize\sf[}}
\newcommand{\closedB}{\makebox[1.4ex][c]{\footnotesize\sf]}}

\newcommand{\sfx}{\text{\smaller\sf X}}
\newcommand{\sfy}{\text{\smaller\sf Y}}
\newcommand{\sfz}{\text{\smaller\sf Z}}
\newcommand{\sfw}{\text{\smaller\sf W}}

\newcommand*{\qed}{\mbox{}\nolinebreak\hfill~\raisebox{0.77ex}[0ex]{\framebox[1ex][l]{}}}

\newtheorem{theorem}{Theorem}
\newtheorem{definition}{Definition}
\newtheorem{ex}[theorem]{Example}
\newenvironment{example}{\begin{ex}\rm}{\end{ex}}

\newcounter{letter}

\renewcommand{\cal}{\mathcal}
\newcommand{\set}[1]{\mbox{$\{#1\}$}}

\begin{document}
\title{One-Tape Turing Machine Variants\\and Language Recognition\,\footnote{%
This article will appear in the Complexity Theory Column of the September 2015 issue of \href{http://www.sigact.org/SIGACT_News/}{\underline{SIGACT News}}}}
\author{%
\setcounter{footnote}{3}
Giovanni Pighizzini\,\footnote{Partially supported by MIUR under the project PRIN ``Automi e Linguaggi Formali: Aspetti Matematici e Applicativi'', code~H41J12000190001.}
\mbox{}\\
{\normalsize Dipartimento di
    Informatica}\\
{\normalsize Universit\`{a} degli Studi di Milano}\\
{\normalsize via Comelico 39, 20135 Milano, Italy}\\
{\normalsize\tt pighizzini@di.unimi.it}%
}%
\date{}%
\maketitle\thispagestyle{empty}%

\begin{abstract}
\noindent
We present two restricted versions of one-tape Turing machines. Both characterize the class of context-free languages.
In the first version, proposed by Hibbard in 1967 and called \emph{limited automata}, each tape cell
can be rewritten only in the first $d$ visits, for a fixed constant~$d\geq 2$.
Furthermore, for~$d=2$ deterministic limited automata are
equivalent to deterministic pushdown automata, namely they characterize deterministic context-free languages.
Further restricting the possible operations, we consider \emph{strongly limited
automata}. These models still characterize context-free languages. However, the
deterministic version is less powerful than the deterministic version of limited automata. In fact, 
there exist deterministic context-free languages that are not accepted by any
deterministic strongly limited automaton.
\end{abstract}

\section{Introduction}

Despite the 
continued
progress in computer technology, one of the main problems
in designing and implementing computer algorithms remains that of finding a good compromise between the production of
efficient algorithms and programs and the available resources.
For instance, up to the 1980s, computer memories were very small. 
Many times, in order to cope with restricted
space availability, computer programmers were forced to choose data structures that are not efficient from the
point of view of the time required by their manipulation.
Notwithstanding the huge increasing of the memory capacities we achieved in the last 20--30 years, in some situations
space remains a critical resource such as, for instance, when huge graphs need to be manipulated,
or programs have to run on some embedded systems or on portable devices.

This is just an example to emphasize the relevance of investigations on computational models operating
under restrictions, which was also 
one of the first research lines in computability and complexity theory.
In particular, starting from the 1960s, a lot of work has been done to investigate 
the minimal amount of resources needed by a machine  in order to be more powerful
than finite state devices. A natural formalization of these problems can be obtained in the realm of formal
languages, by studying resource requirements for nonregular language recognition.
Classical investigations in this field are related to time and space resources and
produced some ``gap results''. 

\medskip

Let us start by briefly considering time.
First of all, using a simple argument, one can prove that each language which is accepted by a Turing machine in 
sublinear time is regular.\footnote{See, e.g.,~\cite{Pi09}.}
Moreover, it should be clear that each regular language can be accepted in linear time by a
Turing machine which never writes symbols, namely by a (two-way) finite automaton.
However, by considering machines with a separate work tape, it is possible to recognize in linear time even
nonregular languages, such as for instance $\{a^nb^n\mid n\geq 0\}$ or the Dyck language of balanced parentheses.

When we consider \emph{one-tape Turing machines}, namely machines with a single (infinite or
semi-infinite) tape, which initially contains the input and which can be rewritten during the computation to
keep information,\footnote{%
As usual, the cells outside the input portion contain a special blank symbol.} the question of the
minimum amount of space needed to recognize nonregular languages becomes more interesting.
In 1965 Hennie proved that deterministic one-tape Turing machines working in linear space are not more powerful than
finite automata, namely, they can recognize only regular languages~\cite{He65}. Hence, the ability to store data on the
tape is useless if the time is restricted to be linear.\footnote{%
Those machines were called \emph{one-tape off-line Turing machines}, where ``off-line'' means that all input symbols are
available, on the tape, at the beginning of the computation and they can be read several times, if not rewritten. 
In contrast, in the 1960s the term ``on-line'' was used
to indicate machines with a separate read-only one-way input tape: in this case the machine can read each input
symbol only one time. So, if an on-line machine has to use the same input symbol more times, it needs to use space to save it somewhere.}
Actually, this remains true even for some time bounds growing more than linear functions. In fact, as independently proved by
Trakhtenbrot and Hartmanis, machines of this kind can recognize nonregular languages only if their
running times grow at least as $n\log n$. Hence, there is a gap between the time sufficient to recognize regular languages, which is linear, and the time necessary for nonregular language recognition.
Furthermore, examples of nonregular languages accepted in time~$O(n\log n)$ have been provided,
proving that the bound is optimal~\cite{Tr64,Ha68}. 

Concerning the nondeterministic case, in 1986 Wagner and Wechsung provided a counterexample
showing that the $n\log n$ time lower bound cannot hold~\cite[Thm.~7.10]{WW86}. This result was improved in~1991 by Michel, by showing the existence
of NP-complete languages accepted in linear time~\cite{Mi91}. Since there are simple examples of regular
languages requiring linear time, we can conclude that, in the nondeterministic case, we do not have a time gap between
regular and nonregular languages.

However, this depends on the time measure we are considering.
In fact, we can take into account all computations (\strong\ measure),
or we can consider, among all accepting computations on each input belonging to the language, the shortest one (\weak\ measure).
This latter measure is related to an optimistic view of nondeterminism: 
On a given input, when a nondeterministic machine guesses an accepting computation,  it is also able to guess the shortest one.

The abovementioned results by Wagner and Wechsung and by Michel have been proved with respect to the \weak\ measure.
For the \strong\ measure the situation is completely different. In fact, under this measure, the $n\log n$ time
lower bound for the recognition of nonregular languages holds even in the nondeterministic case, as proved in 2010
by Tadaki, Yamakami, and Lin~\cite{TYL10}.
Table~\ref{tb:time} summarizes the above-discussed time lower bounds for the recognition
of nonregular languages by one-tape Turing machines.%
\begin{table}[bt]\footnotesize
\begin{center}
\begin{tabular}{l@{\,\,}r@{}}
\begin{tabular}[b]{|l|}\multicolumn{1}{c}{}\\ \hline
~Deterministic machines~\\ \hline
~Nondeterministic machines~\\ \hline
\end{tabular}&
\begin{tabular}[b]{|c|c|@{}}
\hline {\strong}  & {\weak}\\ \hline\hline
$~~n\log n~~$ &  $~~n\log n~~$ \\ \hline
$~~n\log n~~$ &  $~~\hfill n~\hfill$\\ \hline
\end{tabular}
\end{tabular}
\caption{Time lower bounds for the recognition of nonregular languages by one-tape Turing machines.
The table should be
read as follows: a row $r$ denotes a type of machine while
a column $c$ a measure. If the element at the position $(r,c)$ 
of the table is the function $f(n)$, then $t(n)\notin o(f(n))$
for each one-tape off-line 
Turing machine of type $r$ that recognizes a nonregular language
in time $t(n)$ under the measure corresponding to column $c$. 
All the bounds have been proved to be optimal.
For a survey see~\cite{Pi09}.
}
\label{tb:time}
\end{center}
\end{table}

\medskip

Now, let us discuss space. 
It is easy to observe that Turing machines working in constant space can be simulated
by finite automata, so they accept only regular languages.
In their pioneering papers, Hartmanis, 
Stearns, 
and Lewis investigated the minimal 
amount of space that a deterministic Turing machine needs to recognize a
nonregular language~\cite{LSH65, SHL65}.
In order to compare machines with finite automata, counting only the extra space used to keep information,
the machine model they considered has a work tape, which is 
separate from 
the read-only input tape. 
The space is measured only on the work tape.

They proved that if the input tape is \emph{one-way}, namely the input
head is never moved to the left, then, in order to recognize a nonregular language, a logarithmic
amount of space is necessary.
Hence, there are no languages with nonconstant and sublogarithmic space complexity on one-way Turing machines.

In the case of \emph{two-way} machines, namely when machines can move the
head on the input tape in both directions, the lower bound reduces to a double
logarithmic function, namely a function growing as $\log\log n$.\footnote{%
Actually, in these papers the authors used the abovementioned terms \emph{on-line} and \emph{off-line}
instead of \emph{one-way} and \emph{two-way}.}

These results have been generalized to nondeterministic machines by Hopcroft and Ullman
under the \strong\ space measure, namely by taking into account all computations~\cite{HU69}.
The optimal space lower bound for nonregular acceptance on one-way nondeterministic machines reduces to~$\log\log n$,
if on each accepted input the computation using 
least space is considered (\weak\ space), as proved by Alberts~\cite{Al85}.
For a survey on these lower bounds we 
point the reader to~\cite{Me08}.\footnote{%
Many interesting results concerning
``low'' space complexity have been proved. For surveys, see, e.g., the monograph by Szepietowski~\cite{Sz94}
and the papers by Michel~\cite{Mi92} and Geffert~\cite{Ge98}.}

\medskip

Let us now consider \emph{linear space}.
It is well known that nondeterministic Turing machines working within this space bound
characterize the class of context-sensitive languages. This remains true in the
case of \emph{linear bounded automata}, namely one-tape Turing machines whose work space
is restricted to the portion of the tape which at the beginning of the computation contains the input,
as proved in 1964 by Kuroda~\cite{Ku64}.

\medskip

An interesting characterization of the class of context-free languages was
obtained by Hibbard in~1967,  considering a restriction of linear bounded automata,
called \emph{scan limited automata} or, simply, \emph{limited automata}~\cite{Hi67}.
A limited automaton is a Turing machine that can rewrite the content of each tape cell
only 
during the first~$d$ visits, for a fixed constant~$d$. 
It has been observed that the 
restriction of using only 
the portion of the tape
which initially contains the input does not reduce the computational power of these models~\cite{PP14}.
Then, limited automata can be seen as a restriction of linear bounded automata while, in turn,
two-way finite automata, which characterize regular languages, can be seen as restrictions of limited automata.
Hence, we have a hierarchy of classes of one-tape Turing machines which corresponds to the Chomsky hierarchy.

\medskip

This paper is devoted to the description of limited automata and one restricted version of them.
In Section~\ref{sec:limited} we  introduce the model, presenting some examples and some properties.
In particular, we  point out that a deterministic family of limited automata characterizes the
class of deterministic context-free languages. We  also discuss descriptional complexity results,
comparing the size of the description of context-free languages by pushdown automata, with the size
of the description by limited automata.

Since we are interested in devices working with restricted resources or restricted operations, in Section~\ref{sec:strongly}
we will consider a machine model with a set of possible operations which further restricts the operations
available on limited automata. This models is called \emph{strongly limited automata}~\cite{Pi15}.
The idea of studying it was inspired by the fact that Dyck languages, namely  languages of well
balanced sequences of brackets, are recognized by some kinds of limited automata that use the capabilities
of these devices in a restricted way. Furthermore, according to the Chomsky-Sch\"utzenberger representation theorem 
for context-free languages~\cite{CS63}, the ``context-free part'' of each context-free language is
represented by a Dyck language.
Indeed, even if they have severe restrictions, strongly limited automata still characterize the
class of context-free languages. However, their deterministic version is weaker than deterministic pushdown automata,
namely it cannot recognize all deterministic context-free languages.
Even for these devices we  discuss descriptional complexity aspects.

In the final section, we briefly mention further models restricting one-tape machines and related to
context-free language recognition.

\section{Limited Automata}
\label{sec:limited}

Given an integer~$d\geq 0$, a \emph{$d$-limited automaton} is a tuple 
$\mathcal{A}=(Q,\Sigma,\Gamma,\delta,q_0,F)$, where:
\begin{itemize}
\item $Q$ is a finite \emph{set of states}.
\item $\Sigma$ and $\Gamma$ are two finite sets of symbols, called respectively the \emph{input alphabet} and
the \emph{working alphabet}, such that
$\Sigma\cup\set{\leftend,\rightend}\subseteq\Gamma$, where $\leftend$, $\rightend \notin \Sigma$
are two special symbols, called the \emph{left} and the \emph{right end-markers}. 
\item $\delta:Q\times\Gamma\rightarrow 2^{Q\times(\Gamma\setminus\{\leftend,\rightend\})\times\{-1,+1\}}$ 
is the transition function.
\item $q_0\in Q$ is the \emph{initial state}.
\item $F\subseteq Q$ is the \emph{set of final states}.
\end{itemize}

\noindent
At the beginning of the computation, the input is stored onto the tape surrounded by the two end-markers, 
the left end-marker being at position zero. Hence, on input~$w$, the right end-marker is on the cell
in position $|w|+1$. The head of the automaton is on cell~$1$ and the state of the finite control
is~$q_0$.
In one move, according to~$\delta$ and to the current state, $\mathcal{A}$~reads a symbol from the tape,
changes its state, replaces the symbol just read from the tape by a new symbol, 
and moves its head to one position forward or backward.
In particular, $(q,X,m)\in\delta(p,a)$ means that when the automaton in the state~$p$ is
scanning a cell containing the symbol~$a$, it can enter the state~$q$, rewrite the cell content by~$X$,
and move the head to \emph{left}, if $m=-1$, or to \emph{right}, if $m=+1$.
Furthermore, the head cannot violate the end-markers, except at the end of computation,
to accept the input, as explained below.
However, replacing symbols is subject to some restrictions, which, essentially,
allow the modification of the content of a cell only during the first~$d$ visits.
To this aim, the alphabet $\Gamma$ is partitioned into $d+1$ sets $\Gamma_0,\Gamma_1,\ldots,\Gamma_d$,
where $\Gamma_0=\Sigma$ and $\leftend,\rightend\in\Gamma_d$.
With the exception of the cells containing the end-markers, which are never modified,
at the beginning all the cells contain symbols from $\Gamma_0=\Sigma$.
In the $k$-th visit to a tape cell, the content of the cell is rewritten by a symbol from $\Gamma_k$,
up to $k=d$, when the content of the cell is ``frozen'', i.e.,
after that, the symbol in the cell cannot be changed further. Actually, on a cell we do not count the
visits, but the scans from left to right (corresponding to odd numbered visits) and from right to left 
(corresponding to even numbered visits). 
Hence, a move reversing the head direction is counted as a double visit for the cell 
where it occurs. In this way, when a cell $c$
is visited for the $k$th time, with $k\leq d$, its content is a symbol from $\Gamma_{k-1}$.
If the move does not reverse the head direction, then the content of the cell is replaced
by a symbol from $\Gamma_k$. However, if the head direction is reversed, then in this double
visit the symbol is replaced by a symbol from $\Gamma_{k+1}$, when $k<d$, and by a symbol
of $\Gamma_d$ that after then is frozen, otherwise.

Formally, for each $(q,\gamma,m)\in\delta(p,\sigma)$, with $p,q\in Q$, $\sigma\in\Gamma_k$,
$\gamma\in\Gamma_h$, $m\in\{-1,+1\}$, we require the following:
\begin{itemize}
\item if $k=d$ then $\sigma=\gamma$ and $k=h$,
\item if $k<d$ and $m=+1$ then~$h=\min(\lceil\frac{k}{2}\rceil\cdot 2+1,d)$,
\item if $k<d$ and $m=-1$ then~$h=\min(\lceil\frac{k+1}{2}\rceil\cdot 2,d)$.
\end{itemize}
An automaton~$\mathcal{A}$ is said to be \emph{limited} if it is $d$-limited for some $d\geq 0$.
$\mathcal{A}$ accepts an input $w$ if and only if there is a computation path
which starts {}from the initial state~$q_0$ with the input tape containing~$w$ surrounded by the
two end-markers and the head on the first input cell, and which ends in a final state~$q\in F$ after
violating the right end-marker.
The language accepted by~$\mathcal{A}$ is denoted by~$L(\mathcal{A})$.
$\mathcal{A}$~is said to be \emph{deterministic}
whenever $\#{\delta(q,\sigma)}\le 1$, for any $q\in Q$ and
$\sigma\in\Gamma$\@.

\begin{example}
\label{ex:dyck}
For each integer~$k\geq 1$, we denote by~$\Omega_k$ the alphabet of~$k$ types of brackets,
which will be represented as~$\{\openA_1,\closedA_1,\openA_2,\closedA_2,\ldots,\openA_k,\closedA_k\}$.
The \emph{Dyck language}~$D_k$ over the alphabet~$\Omega_k$ is the set of strings representing well 
balanced sequences of brackets.
We will refer to the $\openA_i$ symbols as ``open brackets'' 
and the $\closedA_i$ symbols as ``closed brackets'', i.e. opening
and closing brackets.

The \emph{Dyck language}~$D_k$ can be recognized by a machine~$M$ which
starts having the input string on its tape, surrounded by two end-markers~$\leftend$ and~$\rightend$, 
with the head on the first input symbol.
From this configuration, $M$~moves to the right to find a closed bracket~$\closedA_i$, $1\leq i\leq k$.
Then~$M$ replaces~$\closedA_i$ with a symbol~$\sfx\notin\Omega_k$ and changes the head direction, moving to the left.
In a similar way, it stops when during this scan it meets for the first time a left
bracket~$\openA_j$. If~$i\neq j$, i.e., the two brackets are not of the same type, then~$M$ rejects.
Otherwise, $M$~writes~$\sfx$ on the cell and changes again the head direction moving to the right.
This procedure is repeated until~$M$ reaches one of the end-markers. (See Figure~\ref{fig:brackets}.)

\begin{itemize}
\item If the left end-marker is reached, then it means that at least one of the right brackets in the input
does not have a matching left bracket. Hence, $M$~rejects.

\item If instead the right end-marker is reached, then~$M$ has to make sure that every left bracket has a
matching right one. In order to do this, it scans the entire tape from the right to the left and if
it finds a left bracket not marked with $\sfx$ then~$M$ rejects.
On the other hand, if~$M$ reaches the left end-marker reading only~$\sfx$s, then it can accept the input.
\end{itemize}

\begin{figure}[tb]
\begin{center}
\includegraphics[scale=0.65]{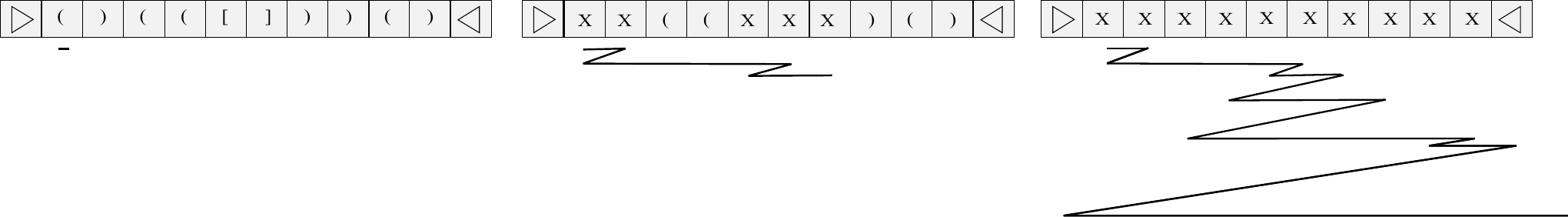}
\end{center}
\vspace*{-10pt}
\caption{%
Some steps in an accepting computation of the automaton~$M$ of Algorithm~\ref{alg:Dyck} on input~
$\protect\openA\protect\closedA\protect\openA\protect\openA\protect\openB\protect\closedB\protect\closedA\protect\closedA\protect\openA\protect\closedA$.
}
\label{fig:brackets}
\end{figure}

\noindent
Now we look more into the details of the implementation of this procedure, which is summarized in
Algorithm~\ref{alg:Dyck}.

\IncMargin{2em}
\begin{algorithm}
  \caption{Recognition of the Dyck language~$D_k$~\label{alg:Dyck}}
\small 
  start with the head on the first input symbol\label{alg:Dyck:start}\;
  \While{symbol under the head $\neq\rightend$}{\label{alg:Dyck:mainLoopBegin}%
    move the head to the right\label{alg:Dyck:moveRight}\;
    \If{symbol under the head $=\,\closedA_i$ \mbox{\rm(with~$1\leq i\leq k$)}\label{alg:Dyck:if}}{%
    	write~$\sfx$\label{alg:Dyck:writeRight}\;
	    \Repeat{\label{alg:Dyck:intUntil}symbol under the head $\neq \sfx$}{\label{alg:Dyck:intRepeat}%
	    move the head to the left\label{alg:Dyck:moveLeft}\;
	    }
	    \lIf{symbol under the head $\neq\,\openA_i$}{{\sc Reject}\label{alg:Dyck:internalReject}}
    	write~$\sfx$\label{alg:Dyck:writeLeft}\;
    } 
  }
  \Repeat{\label{alg:Dyck:extUntil}symbol under the head $\neq \sfx$}{\label{alg:Dyck:extRepeat}%
  move the head to the left\label{alg:Dyck:extMoveLeft}\;
  }
	\lIf{symbol under the head $=\leftend$}{{\sc Accept}\label{alg:Dyck:acc}}
	\lElse{{\sc Reject}\label{alg:Dyck:rej}}
\end{algorithm}

The machine~$A_D$ starts the computation in the initial state~$q_0$ (line~\ref{alg:Dyck:start}).
While moving to the right to search for a closed bracket (line~\ref{alg:Dyck:moveRight}), 
$A_D$~does not need to keep in its finite control any other information, so it can
always use the same state~$q_0$.
On the other hand, after a closed bracket~$\closedA_i$ is found, $1\leq i\leq k$, $A_D$~needs to remember
the index~$i$ in order to find a matching open bracket~$\openA_i$. This is done by using
a state~$q_i$ for moving to the left while performing the search and the next rewriting
(lines~\ref{alg:Dyck:intRepeat}--\ref{alg:Dyck:writeLeft}).
The final loop and test (lines~\ref{alg:Dyck:extRepeat}--\ref{alg:Dyck:rej}) can be performed
using a further state $q_{\rightend}$.
Notice that each cell containing a closed bracket is rewritten in the first visit, while changing the head
direction, and each cell containing an open bracket is rewritten in the second visit.
Furthermore, the content of a cell is not rewritten after the second visit. Hence
the machine~$A_D$ we just described is a deterministic \la{2}.\footnote{%
According to the definition of limited automaton, the alphabet~$\Gamma$ should be partitioned in
three sets $\Gamma_0=\Sigma$, $\Gamma_1$, and $\Gamma_2$, and each open bracket should be rewritten
by a symbol of~$\Gamma_1$ in the first visit. This can be trivially done by replacing the open bracket
with a marked version. However, for the sake of simplicity, here and in the next examples we prefer to avoid
these details.}
\qed
\end{example}

\begin{example}
\label{ex:Kn}
For each integer $n$, let us denote by $K_n$ the set of all strings over the
alphabet~$\{0,1\}$ consisting of the concatenation of blocks of length $n$, 
such that at least $n$ blocks are equal to the last one. Formally:
\begin{eqnarray*}
K_n&=&\{x_1x_2\cdots x_kx\mid k\geq 0,~x_1,x_2,\ldots,x_k,x\in\{0,1\}^n,\\
&&~\exists i_1<i_2<\cdots<i_n\in\{1,\ldots,k\},~x_{i_1}=x_{i_2}=\ldots=x_{i_n}=x\}\,.
\end{eqnarray*}
We now describe a \twol~$M$ accepting ~$K_n$.
Suppose $M$ receives an input string $w$ of length~$N$.
\begin{enumerate}
\item First, $M$ scans the input tape from left to right, to reach the right end-marker.
\item $M$ moves its head $n+1$ positions to the left, namely to the cell~$i=N-n$, the one immediately to the
left of the input suffix~$x$ of length~$n$.
\item Starting from this position~$i$, $M$ counts how many blocks of length~$n$ coincide
	with~$x$. This is done as follows.
	
	When~$M$, arriving from the right, visits a position $i\leq N-n$ for the first time,
	it replaces the content $a$ by a special symbol $X$, after copying~$a$ in the finite control.
	Hence, $M$ starts to move to the right, in order to compare the symbol removed from the cell
	with the corresponding symbol in the block~$x$. While moving to the right,
	$M$~counts modulo~$n$ and stops when the counter is~$0$ and a cell containing a symbol
	other than~$X$ is reached.\footnote{We remind the reader that~$M$ has to recognize
	the language~$K_n$ for a \emph{fixed} integer~$n$.} The symbol of~$x$ in this cell  has to be
	compared with~$a$.
	Then,~$M$ moves to the left until it reaches cell~$i-1$, namely the first cell
	which does not contain~$X$, immediately to the left of cells containing~$X$.
	
	We observe that the end of a block is reached each time a symbol~$a$ copied from the
	tape is compared with the leftmost symbol of~$x$, which 
lies
immediately to the
	right of a cell containing~$X$. If in the block just inspected
	no mismatches have been discovered then	the counter of blocks matching with~$x$ is incremented
	(unless its value was already~$n$).
	
	\item When the left end-marker is reached, $M$ accepts if and only if the input
	length is a multiple of~$n$ and the counter of blocks matching with~$x$ contains~$n$.
\end{enumerate}
We can easily observe that the above strategy can modify tape cells only in the first two
visits. Hence, it can be implemented by a \emph{deterministic} \twol. Such an automaton
uses~$O(n^2)$ states and a constant size alphabet.

\medskip

Actually, using nondeterminism, it is possible to recognize the language~$K_n$ using~$O(n)$  states
and modifying tape cells only in the first visit, namely~$K_n$ is accepted by a nondeterministic \onel~$M$,
which we now describe.\footnote{%
The presentation is adapted from a similar example presented in~\cite{PP14}, where more
details can be found.}  $M$~works in three phases.
\begin{enumerate}
  
\item First, $M$ scans its tape from left to right. During this phase, $M$ marks exactly~$n+1$ input
  cells. The first~$n$ marked cells are guessed to be the leftmost positions of the $n$ input blocks
  $x_{i_1},\ldots,x_{i_n}$ which are expected to be equal to the rightmost block. The last marked cell is guessed to
  be the leftmost position of the rightmost block.  
  This phase can be implemented using~$n+2$ states, to count how many positions have been marked.
  
\item When the right 
end-marker
is reached, $M$~makes a complete scan of the input 
  from right to left, in order to verify whether or not the input length is a multiple of~$n$, the
  last cell that has been marked in the first phase is the leftmost cell of the last block, and the other
  marked cells are the leftmost cells of some blocks. If the outcome of this phase is negative,
  then~$M$ stops and rejects.
  The number of states used here is~$2n$.
  
\item  Finally, $M$ verifies that all the blocks starting from the marked positions
  contain the same string of length~$n$.
  To this aim, the block from position $j_{h-1}$ is compared symbol by symbol with the block
  from position $j_h$, for~$h=1,\ldots,n$, where~$j_0<j_1<\cdots<j_n$ are the marked positions.
  To make these comparisons, $M$ with the head on the $i$th symbol of the block~${h-1}$ moves the head
  to the $i$th symbol of the block~$h$. To detect the position of such symbol, $M$ while moving
  the head to the right counts modulo~$n$, until, after visiting a marked cell (namely cell~$j_h$), 
  the value of the counter becomes~$0$. If the comparison fails, then~$M$ can
  reject, otherwise~$M$ starts to move to the left, still counting modulo~$n$, and searching the leftmost
  cell of the block~$h-1$. Then~$M$ can start to move to the right, while decrementing the counter, reaching
  cell~$i$ when the counter is~$0$ and so locating cell~$i+1$ to start the next comparison. 
  Furthermore, the value of the counter also allows 
the discovery of whether
all~$n$ symbols of the block have been inspected.
  The implementation of this phase uses $O(n)$ states.
 \qed 
\end{enumerate}  
\end{example}

\subsection*{Computational power}
The following theorem summarizes the most important known results on the computational power of
$d$-limited automata:

\begin{theorem}
\label{th:powerlimited}
\begin{enumerate}
\item[(i)]For each $d\geq 0$, the class of languages accepted by \las{d} coincides with the class of context-free languages~\cite{Hi67}.
\item[(ii)]The class of languages accepted by deterministic \las{2} coincides with the class of deterministic context-free languages~\cite{PP15}.
\item[(iii)]For each $d\geq 2$, there exists a language which is accepted by a deterministic \la{d},
which cannot be accepted by any deterministic \la{(d-1)}~\cite{Hi67}.
\item[(iv)]The class of languages accepted by \las{1} coincides with the class of regular languages~\cite{WW86}.
\end{enumerate}
\end{theorem}

The argument used by Hibbard to prove Theorem~\ref{th:powerlimited}(i) is very difficult.
He  provided some constructions to transform  
a kind of rewriting system, 
equivalent to
pushdown automata, to \las{2} and vice versa, together with reductions from \las{(d+1)} to
\las{d}, for~$d\geq 2$~\cite{Hi67}.

We now discuss a construction of \las{2} from context-free languages (already presented in~\cite{PP14}), which is based on 
the Chomsky-Sch\"utzenberger representation theorem for context-free languages~\cite{CS63}.
We remind the reader that this theorem states that each context-free language can be obtained by selecting in a Dyck language~$D_k$, with~$k$
kinds of brackets, only the strings belonging to a regular language~$R$, and then renaming the symbols in the remaining strings
according to a homomorphism~$L$.
More precisely, every context-free language~$L\subseteq\Sigma^*$ can be expressed
as~$L=h(D_k\cap R)$, where $D_k\subseteq\Omega_k^*$, $k\geq 1$, is a Dyck language,
$R\subseteq\Omega_k^*$ is a regular language, and~$h:\Omega_k\rightarrow\Sigma^*$ is a homomorphism.

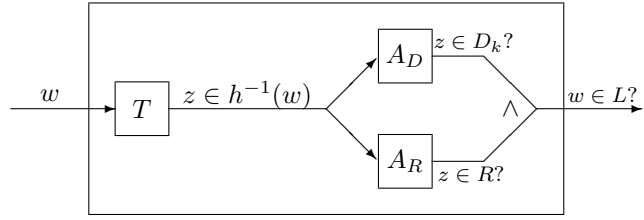
\begin{figure}
	\setlength{\unitlength}{0.35cm}	
	\begin{center}
	\begin{picture}(24,8)(-1,0)
		\put(3,3){\framebox(2,2){\small $T$}}
    \put(13,1){\framebox(2,2){\small $A_R$}}
		\put(13,5){\framebox(2,2){\small $A_D$}}
		\put(2,0){\framebox(18,8){}}
		
		\put(-1,4){\vector(1,0){4}}
		\put(0.5,4.5){\makebox(0,0){\small $w$}}
		
		\put(5,4){\line(1,0){6}}
		\put(8,4.5){\makebox(0,0){\small $z\in h^{-1}(w)$}}
		\put(11,4){\vector(1,1){1.9}}
		\put(11,4){\vector(1,-1){1.9}}
%
		\put(15,2){\line(1,0){2}}
		\put(16.5,1.5){\makebox(0,0){\scriptsize $z\in R$?}}
		\put(17,6){\line(1,-1){2}}
%
		\put(15,6){\line(1,0){2}}
		\put(16.6,6.5){\makebox(0,0){\scriptsize $z\in D_k$?}}
		\put(17,2){\line(1,1){2}}

		\put(19,4){\vector(1,0){4}}
		\put(21.5,4.5){\makebox(0,0){\scriptsize $w\in L$?}}
		\put(18,4){\makebox(0,0){\small $\wedge$}}	
	\end{picture}
	\caption{A machine accepting $L=h(D_k\cap R)$\label{fig:cs}}
	\end{center}
\end{figure}

Hence, given~$L\subseteq\Sigma^*$ context-free, we can consider the following machines:
\begin{itemize}
\item A  nondeterministic transducer~$T$ computing~$h^{-1}$.
\item The \la{2}~$A_D$ described in Example~\ref{ex:dyck} recognizing the Dyck language $D_k$.
\item A finite automaton~$A_R$ accepting the regular language~$R$.
\end{itemize}
To decide if a string~$w\in\Sigma^*$, we can combine these machines as in Figure~\ref{fig:cs}.

Now, we discuss how to embed the transducer~$T$, the \la{2}~$A_D$, and the automaton~$A_R$ 
in a unique \la{2}~$M$, \emph{provided that the homomorphism~$h$ is non-erasing}, namely $h(\sigma)\neq\epsilon$, for each $\sigma\in\Omega_k$.

In a first phase, $T$ and~$A_D$ work together using a producer--consumer scheme and, after that,
in a second phase~$A_R$ is simulated. In the first phase, when $A_D$ has to examine for the first
time a tape cell, $T$ produces in a nondeterministic way a symbol $\sigma\in h^{-1}(u)$, 
for a nondeterministically chosen prefix~$u$ of the part of the input~$w$ which starts from 
$T$'s
current head position.
Then, a move of~$A_D$ which rewrites~$\sigma$ by a new symbol~$\sigma'$ is simulated.
The symbol~$\sigma'$ needs to be stored somewhere in the case~$A_D$
has to visit the same tape cell during the computation.
Furthermore, the symbol~$\sigma$ will be used for the simulation of $A_R$. Hence, the machine~$M$ has to keep both symbols~$\sigma$ and~$\sigma'$.
Since~$h$ is non-erasing, this can be done by replacing~$u$ by the pair $(\sigma, \sigma')$ on the tape. 
More precisely, the tape of~\la{2}~$M$ is divided in two tracks. 
At the beginning of the computation, the first track contains the input~$w$,
while the second track is empty. In the first phase, the finite control of~$M$ simulates the 
controls of both~$T$ and~$A_D$. 
$M$~alternates the simulation of some computation steps of~$T$ with the simulation of some computation steps of~$A_D$ as follows:
\begin{enumerate}
\item\label{step1} When the head reaches a cell which has not yet been visited (hence, also at the beginning of
  the computation),~$M$ simulates~$T$, by nondeterministically replacing a prefix~$u$ of the remaining input, 
  with a string $\sharp^{|u|-1}\sigma$ such that $\sigma\in h^{-1}(u)$. The symbol~$\sharp$ is used for padding
  purposes. All the cells containing this symbol will be skipped in the future steps.
  
\item\label{step2} In the last step of the above-described part of computation, when the rightmost symbol of~$u$ is replaced 
  by~$\sigma$ on the first track, $M$ also resumes the simulation of~$A_D$, starting from a step 
  reading~$\sigma$. Hence, while writing~$\sigma$ on the first track, $M$~also writes on the second track
  the symbol $\sigma'$ which is produced by~$A_D$, while rewriting~$\sigma$ in the first visit.
\item If~$A_D$ moves to the left, going back to already-visited cells, then $M$~simulates directly the moves of~$A_D$,
  skipping all cells containing~$\sharp$, and using the second track of the tape.
  When, moving to the right, the head of~$M$ reaches a cell which has not been
  visited before, the simulation of~$A_D$ is interrupted. (A cell not visited before can be located
  since it does not contain $\sharp$ and the second track is empty.) In this case, $M$ resumes the
  simulation of~$T$, as  explained at point~\ref{step1}, except in the case the cell contains the right end-marker.
\item When the right end-marker is reached, the first track contains a string 
  $z\in h^{-1}(w)$, while the second track contains the result of the 
  rewriting of~$z$ by~$A_D$ (ignoring all the cells containing~$\sharp$).
  If $A_D$ rejects, namely, $z\notin D_k$, then $M$ rejects. Otherwise, $M$ moves its head to the left end-marker
  and, starting from the first tape cell, it simulates the automaton~$A_R$, consulting the first track, in order to
  decide whether or not $z\in R$. Finally, $M$ accepts if and only if $A_R$ accepts.
\end{enumerate}
Actually, the simulation of the automaton~$A_R$ can be done in the first phase, while simulating~$T$ 
and~$A_D$. In particular, at the previous point~\ref{step2}, when~$T$ produces a symbol~$\sigma$ and $M$ 
simulates a move of~$A_D$ on~$\sigma$, $M$ can also simulate a move of~$A_R$ on~$\sigma$. To this aim,
$M$ has to keep in its finite state control, together with the controls of~$T$ and~$A_D$, also the control
of~$A_R$. This increases the number of the states of~$M$, but makes superfluous having the first track to keep the
string~$z$. Hence, it reduces the size of the working alphabet of~$M$.

In this construction, we used the hypothesis that the homomorphism~$h$ is non-erasing.
Actually, the classical proof of the Chomsky-Sch\"utzenberger representation theorem produces an erasing homomorphism.
However, an interesting variant of the theorem, recently obtained by Okhotin, states that for
$L\subseteq\Sigma^*\setminus\Sigma$ we can always restrict ourselves to \emph{non-erasing} homomorphisms, namely,
we can consider $h:\Omega_k\rightarrow\Sigma^+$~\cite{Okh12}.\footnote{The restriction 
$L\subseteq\Sigma^*\setminus\Sigma$ can be easily removed after observing that one-letter strings can be trivially recognized
without any rewriting.}

\medskip

We point out that the construction of \las{2} we just outlined always produces a nondeterministic machine,
since the transducer~$T$ is intrinsically nondeterministic.

Actually, the transformation from pushdown automata to \la{2} provided by Hibbard preserves the 
determinism. Hence, each deterministic context-free language is accepted by a deterministic \la{2}.
The converse inclusion, which was left open by Hibbard,  has been recently proved, thus obtaining
that the class of languages accepted by deterministic \las{2} coincides with the class of deterministic
context-free languages (Theorem~\ref{th:powerlimited}(ii))~\cite{PP15}.

It is natural to ask what is the power of determinism in the case of \las{d}, with~$d\geq 3$.
It is not difficult to describe a deterministic \la{3} accepting the nondeterministic context-free language
$\{a^nb^nc\mid n\geq 0\}\cup\{a^nb^{2n}d\mid n\geq 0\}$. Indeed, Hibbard showed the existence of an infinite
deterministic hierarchy (Theorem~\ref{th:powerlimited}(iii)).\footnote{Hibbard proved the separations in 
Theorem~\ref{th:powerlimited}(iii) for~$d\geq 3$.
In the case~$d=2$ the separation follows from the fact that 
the \las{1} accept 
only regular languages.}

\medskip

The  simulation of \las{1} by finite automata  (Theorem~\ref{th:powerlimited}(iv)) has been obtained in~\cite[Thm.\ 12.1]{WW86} by
adapting standard techniques involving transition tables. These techniques have been introduced for converting two-way automata into equivalent
one-way automata~\cite{Sh59}.

\subsection*{Descriptional power}

Descriptional complexity aspects related to the results in Theorem~\ref{th:powerlimited} have been recently investigated
providing new conversions between \las{2} and pushdown automata and between \las{1} and finite automata.

\begin{theorem}[\cite{PP15}]\label{th:costspda}~
\begin{itemize}
\item[(i)]Each $n$-state \la{2} can be simulated by a pushdown automaton of size exponential in a polynomial in~$n$.
\item[(ii)]The previous upper bound becomes a double exponential when a deterministic \la{2} is simulated by a deterministic pushdown automaton,
  however it remains a single exponential if the input of the deterministic pushdown automaton is given with a symbol to mark the right end.
\item[(iii)]Each pushdown automaton~$M$ can be simulated by a \la{2} whose size is polynomial with respect to the size of~$M$.
\item[(iv)]The previous upper bound remains polynomial when a deterministic pushdown automaton is simulated by a deterministic \la{2}.
\end{itemize}
\end{theorem}

\noindent
The exponential gap for the conversion of \las{2} into equivalent pushdown automata cannot be reduced. In fact, the language~$K_n$
presented in Example~\ref{ex:Kn} is accepted by a (deterministic) \la{2} with~$O(n^2)$ states and a constant size alphabet, while
the size of each pushdown automaton accepting it must be at least exponential in~$n$~\cite{PP15}. This also implies that the
simulation of deterministic \las{2} by deterministic pushdown automata is exponential in size. Actually, we conjecture that
this simulation costs a double exponential, namely it matches the upper bound in Theorem~\ref{th:costspda}(ii).

\begin{theorem}[\cite{PP14}]
\label{th:reg}
  Each~$n$-state \onel~$M$ can be simulated by a nondeterministic automaton
	with~$n\cdot 2^{n^2}$ states and by a deterministic automaton with $2^{n\cdot 2^{n^2}}$ states.
	Furthermore, if~$M$ is deterministic then an equivalent \owdfa\ with no more
	than~$n\cdot(n+1)^n$ states can be obtained.
\end{theorem}

The doubly exponential upper bound for the conversion of nondeterministic \las{1} into deterministic automata
is related to a double role of nondeterminism in \las{1}. 
When a  \onel\ visits one cell after the first rewriting, the possible nondeterministic transitions depend on the
symbol that has been written in the cell in the first visit, which, in turns, depends on the nondeterministic choice taken in the first visit.
This double exponential cannot be avoided. In fact, as we already observed, the language $K_n$ of Example~\ref{ex:Kn}
is accepted by a nondeterministic \la{1} with~$O(n)$ states and, using standard distinguishability arguments, 
it can be shown that each deterministic automaton accepting it requires
a number of states doubly exponential in~$n$.
As observed in~\cite{PP14}, even the simulation of deterministic \las{1} by \emph{two-way nondeterministic automata}
is exponential in size.

\medskip

It should be interesting to investigate the size costs of the simulations of \las{d} by pushdown automata for~$d\geq 0$.

\medskip

We conclude this section by briefly mentioning the case of unary languages.
It is well-known that in this case each context-free language is regular~\cite{GR62}.
Hence, for each~$d\geq 0$, \las{d} with a one letter input alphabet recognize only regular languages.
In~\cite{PP14}, a result comparing the size of unary \las{1} 
with the size of equivalent two-way nondeterministic finite automata has been obtained. Quite recently, Kutrib and Wendlandt proved
state lower bounds for the simulation of unary \las{d} by different variants of finite automata~\cite{KW15}.

\section{Strongly Limited Automata}
\label{sec:strongly}

In Section~\ref{sec:limited}, using the Chomsky-Sch\"utzenberger representation theorem for context-free
languages, we discussed how to construct a \la{2} accepting a given context-free language.
The main component of such an automaton is  a \la{2} accepting a Dyck language~$D_k$. However, Algorithm~\ref{alg:Dyck} described in
Example~\ref{ex:dyck} recognizes~$D_k$ without fully using the capabilities of
2-limited automata. For instance, it does not need to rewrite each tape cell two times.
So, we can ask if it is further possible to restrict the moves of 2-limited automata, still keeping
the same computational power.

In~\cite{Pi15} we gave a positive answer to this question, by introducing \emph{strongly limited automata},
a restriction of limited automata which closely imitates the moves which are used in Algorithm~\ref{alg:Dyck}.
In particular, these machines satisfy the following restrictions:
\begin{itemize}
\item While moving to the right, a strongly limited automaton always uses the same state~$q_0$ until the content of a cell
(which has not been yet rewritten) is modified. Then it starts to move to the left.
\item While moving to the left, the automaton rewrites each cell it meets that is not yet rewritten up to some position
where it starts again to move to the right. Furthermore, while moving to the left the automaton does not change its
internal state.
\item In the final phase of the computation, the automaton inspects all tape cells, to check whether or not the final
content belongs to a given 2-strictly locally testable language. Roughly, this means that all the factors of two letters
of the string which is finally written on the tape\footnote{The string which is considered \emph{includes the end-markers.}} should belong to a given set.
\end{itemize}
We now present the detailed definition of this model and then we discuss it.

\begin{definition}
\label{def:sla}
	A \emph{strongly limited automaton} is a tuple ${\cal M}=(Q,\Sigma,\Gamma,\delta,q_0,q_{\leftend})$,
	where:
	\begin{itemize}
	\item $Q$~is a finite \emph{set of states}, which is partitioned in the three disjoint sets
		$\{q_0\}$, $Q_L$, and~$Q_{\Upsilon}$.
	\item $\Sigma$ and $\Gamma$ are two finite and disjoint sets of symbols, called respectively
	  the \emph{input alphabet} and the \emph{working alphabet} of~${\cal M}$.
	  Let us denote by~$\Upsilon$ the \emph{global alphabet} of~${\cal M}$ defined as
	  $\Upsilon=\Sigma\cup\Gamma\cup\{\leftend,\rightend\}$, where $\leftend,\rightend\notin\Sigma\cup\Gamma$
	  are, respectively, the \emph{left} and the \emph{right end-marker}.
	\item $\delta:Q\times\Upsilon\rightarrow 2^{\mbox{\scalebox{0.76}{$\{\moveL,\moveR,\moveWL{\sfx},\turnL{\sfx}{q},\turnR{\sfx}{q},\checkL{q\,}\mid
	\sfx\in\Gamma, q\in Q\}$}}}$ is the \emph{transition function}, which associates a set of possible \emph{operations} with each configuration of~${\cal M}$.
	\item $q_0$ is the \emph{initial state}.
	\item $q_{\leftend}\in Q_{\Upsilon}$ is the \emph{final state}.
	\end{itemize}
The transition function $\delta$ has to satisfy the conditions listed below.
	\begin{itemize}
	\item For the state~$q_0$:
		\begin{itemize}
		\item $\delta(q_0,a)=\{\moveR\}$ if $a\in\Gamma$,
		\item $\delta(q_0,a)\subseteq\{\moveR\}\cup\{\turnL{\sfx}{q}\mid q\in Q_L, \sfx\in\Gamma\}$ if $a\in\Sigma$,
		\item $\delta(q_0,\rightend)=\{\checkL{q_{\rightend}}\}$,
		\item $\delta(q_0,\leftend)$ is undefined.
		\end{itemize}
	\item For each state~$q\in Q_L$:
		\begin{itemize}
		\item $\delta(q,a)=\{\moveL\}$ if $a\in\Gamma$,
		\item $\delta(q,a)\subseteq\{\moveWL{\sfx},\turnR{\sfx}{q_0}\mid \sfx\in\Gamma\}$ if $a\in\Sigma$,
		\item $\delta(q,a)$ is undefined if $a\in\{\leftend,\rightend\}$.
		\end{itemize}
	\item For each state~$q_\sfx\in Q_\Upsilon$:
		\begin{itemize}
		\item $\delta(q_\sfx,a)\subseteq\{\checkL{q_a}\}$, where $a\in\Upsilon$. 
		\end{itemize}
	\end{itemize}
\end{definition}
We now describe how~$\cal M$ works, providing an informal explanation of the meaning of the states and 
of the operations that~$\cal M$ can perform.
First of all, we assume that at the beginning of the computation the tape contains the input string
$w\in\Sigma^*$ surrounded by the two end-markers. Tape cells are counted from~$0$. Hence,
cell~$0$ contains $\leftend$ and cell~$|w|+1$ contains~$\rightend$.
The head is on cell~$1$, namely scanning the leftmost symbol of~$w$, while the finite control is in~$q_0$.

The initial state $q_0$ is the only state which is used while moving from left to right.
In this state all the cells that have been already rewritten are ignored, just moving one position
further, while on all the other cells~$\cal M$ could be allowed either to move to the right or to rewrite the
cell content and then turn the head direction to the left, entering a state in the set~$Q_L$.
To this aim, in the state~$q_0$ the following operations could be possible:
\begin{itemize}
	\item\emph{Move to the right} $\moveR$\\
	Move the head one position to the right \emph{without} rewriting the cell content and \emph{without} 
	changing the state.
	\item\emph{Turn to the left} $\turnL{\sfx}{q}$\\
	Write~$\sfx\in\Gamma$ in the currently scanned tape cell, move one position to the left, entering in 
	state~$q\in Q_L$.
	After a sequence of moves from left to right,
	with this operation $\cal M$ rewrites the content of the current cell and changes the head direction, entering a state~$q\in Q_L$.
\end{itemize}
We point out that these two operations are not allowed in states other than~$q_0$.
One further operation ($\checkL{q_{\rightend}}$, described later) is possible in~$q_0$, 
when the right end-marker is reached, to activate the final phase of the computation.

The states in the set~$Q_L$ are used to move to the left. In a state~$q\in Q_L$, the automaton~$\cal M$
ignores all the cells that have been already rewritten, just moving to the left. On the remaining cells
that $\cal M$~visits, it always rewrites the content up to some position where it turns its head to the right.
During this procedure, $\cal M$~changes state only at the end, when it enters again in~$q_0$.
In a state $q\in Q_L$ the following operations can be allowed:
\begin{itemize}
	\item\emph{Move to the left} $\moveL$\\
	Move the head one position to the left \emph{without} rewriting the cell content and \emph{without} 
	changing the state. This move is used only on cells that have been rewritten.

	\item\emph{Write and move to the left} $\moveWL{\sfx}$\\
	Write~$\sfx\in\Gamma$ in the currently scanned tape cell, move one position to the left, \emph{without} changing the state.
	This move can be used only on cells not yet rewritten.
	
	\item\emph{Turn to the right} $\turnR{\sfx}{q_0}$\\
	Write~$\sfx\in\Gamma$ in the currently scanned tape cell, move one position to the right, entering in 
	the state~$q_0$. Even this move can be used only on cells not yet rewritten.
\end{itemize}
If the left end-marker is reached while scanning to the left in a state of~$Q_L$ then the computation stops
by rejecting (technically the next transition is undefined).
On the other hand, if the right end-marker is reached while scanning to the right in~$q_0$, the machine
starts a final phase where it completely scans the tape from right to left and then stops.
During the last phase~$\cal M$ checks the membership of the final tape content to a \emph{local language}.\footnote{%
A regular language~$L$ is said to be \emph{strictly locally testable} 
if there is an integer $k$ such that membership of a string~$x$ to~$L$ can be ``locally''
verified by inspecting all factors of length~$k$ in~$x$~\cite{MP71}. 
In the case~$k=2$ we simply say that the language is \emph{local}.
More precisely, given an alphabet $\Delta$ and two extra symbols $\leftend,\rightend\notin\Delta$,
we say that a language $L\subseteq\Delta^*$ is \emph{local} if and only if
there exists a set~$F\subseteq(\Delta\cup\{\leftend,\rightend\})^2$ of \emph{forbidden
factors} such that a string $x\in\Delta^*$ belongs to~$L$ if and only if no factor of
length~$2$ of $\leftend x\rightend$ belongs to~$F$.
}
If some forbidden factor is detected then the next transition is undefined and hence~$\cal M$ rejects.
To this aim, in this phase only states from the set~$Q_\Upsilon$ are used.
We assume that there is a surjective map from~$\Upsilon$ to~$Q_\Upsilon$.
We simply denote as~$q_\sfx$ the state associated with the symbol~$\sfx\in\Upsilon$. Note that~$\sfx\neq \sfy$ does
not implies $q_\sfx\neq q_\sfy$.
The following operation is used in this phase:
\begin{itemize}
	\item\emph{Check to the left} $\checkL{q_a}$\\
	On a cell containing symbol $a\in\Upsilon$, move to the left remembering the state associated with~$a$.
\end{itemize}
If no forbidden factor is found, $\cal M$~finally violates the
left end-marker in the state~$q_{\leftend}$. In this case the input is accepted.
Otherwise the computation of~$\cal M$ stopped in some previous step, rejecting the input.
Hence, we assume that~\emph{$\cal M$ accepts its input if and only if from the cell containing
the left end-marker it can further move to the left entering the final state~$q_{\leftend}$.}

\begin{example}
\label{ex:sla}
    Consider the alphabet $\Omega_2$, with brackets represented by the symbols~$\openA,\closedA,\openB,\closedB$.
    The Dyck language~$D_2$ is accepted by a strongly limited automaton with $\Gamma=\{\sfx\}$, $Q_L=\{q_1,q_2\}$, 
    $Q_\Upsilon=\{q_\sfx,q_{\leftend},q_{\rightend}\}$, and the following transitions (we omit braces
    for the sake of the brevity):
    \begin{itemize}
    \item $\delta(q_0,\openA)=\delta(q_0,\openB)=\,\moveR$,
          $\delta(q_0,\closedA)=\,\turnL{\sfx}{q_1}$, $\delta(q_0,\closedB)=\,\turnL{\sfx}{q_2}$,
    \item $\delta(q_1,\sfx)=\delta(q_2,\sfx)=\,\moveL$,
          $\delta(q_1,\openA)=\delta(q_2,\openB)=\,\turnR{\sfx}{q_0}$,
    \item $\delta(q_0,\rightend)=\,\checkL{q_{\rightend}}$, 
    	$\delta(q_{\rightend},\sfx)=\delta(q_\sfx,\sfx)=\,\checkL{q_{\sfx}}$,
    	$\delta(q_\sfx,\leftend)=\,\checkL{q_{\leftend}}$.
    \end{itemize}
    It can be observed that the states $q_{\rightend},q_\sfx,q_{\leftend}$, used in the final scan,
    can be merged in a unique state~$q_{\leftend}$. In fact, in this example the purpose of the final
    scan is to check that all the input symbols have been rewritten, namely, no
    symbol $a\in\{\openA,\closedA,\openB,\closedB\}$ is left on the tape. If such a symbol is discovered,
    then the next transition is not defined and hence the computation rejects.
\qed
\end{example}

\begin{example}
\label{ex:anb12n}
  The deterministic context-free language $\{a^nb^{2n}\mid n\geq 0\}$ is accepted by a strongly limited automaton which guesses each second~$b$.
  While moving from left to right and reading~$b$, the automaton makes a nondeterministic choice between further moving to the right 
  or rewriting the cell by~$\sfx$ and turning to the left.
  Furthermore, while moving to the left, the content of each cell containing~$b$ which is visited is rewritten by~$\sfy$, still
  moving to the left, and when a cell containing~$a$ is visited, its content is replaced by $\sfz$, turning to the right.
  In the final scan the machine accepts if and only if the string on the tape belongs to $\leftend\sfz^*(\sfy\sfx)^*\rightend$.  
  
  We can modify the above algorithm to recognize the language~$\{a^nb^n\mid n\geq 0\}\cup\{a^nb^{2n}\mid n\geq 0\}$.
  While moving from left to right, when the head reaches a cell containing~$b$ three actions are possible: either the
  automaton continues to move to the right, without any rewriting, or it rewrites the cell by $\sfz$, turning to the right,
  or it rewrites the cell by $\sfw$, also turning to the right. While moving from right to left, the automaton behaves as
  the one above described for~$\{a^nb^{2n}\mid n\geq 0\}$.
  The input is accepted if and only if the string which is finally on the tape 
  belongs to~$\leftend\sfz^*\sfw^*\rightend+\leftend\sfz^*(\sfy\sfx)^*\rightend$.
  \qed
\end{example}

We already mentioned that strongly limited automata have the same computational power as limited automata, namely they characterize
context-free languages. This result has been proved in~\cite{Pi15}, also studying descriptional complexity aspects:

\begin{theorem}[\cite{Pi15}]
\label{th:sla}~
\begin{itemize}
\item[(i)]Each context-free language~$L$ is accepted by a strongly limited automaton whose description has a size
  which is polynomial with respect to the size of a given context-free grammar generating~$L$ or of a given
	pushdown automaton accepting~$L$.
\item[(ii)]Each strongly limited automaton~$\cal M$ can be simulated by a pushdown automaton of size polynomial with
  respect the size of~$\cal M$.
\end{itemize}
\end{theorem}
The proof of~(i) was obtained using a further variant of the Chomsky-Sch\"utzenberger representation theorem, also proved
by Okhotin~\cite{Okh12}. In this variant Dyck languages extended with neutral symbols and letter-to-letter
homomorphisms are used.
(ii)~has been proved by providing a direct simulation.

\medskip

Concerning deterministic computations, it is not difficult to observe that deterministic strongly limited automata
cannot recognize all deterministic context-free languages.
Consider, for instance, the deterministic language~$L=\{ca^nb^n\mid n\geq 0\}\cup\{da^{2n}b^n\mid n\geq 0\}$.
While moving from left to right, a strongly limited automaton can use only the state~$q_0$. Hence, it cannot
remember if the first symbol of the input is a~$c$ or a~$d$ and, then, if it has to check whether the number
of~$a$s is equal to the number of~$b$s or whether the number of~$a$s is two times the number of~$b$s.
A formal proof that the language~$L$, and also the language~$\{a^nb^{2n}\mid n\geq 0\}$ (Example~\ref{ex:anb12n}),
are not accepted by any deterministic strongly limited automaton is presented in~\cite{Pi15}.

In that paper it was also proposed to slightly relax the definition of strongly limited automata, 
by introducing a set of states~$Q_R$, with~$q_0\in Q_R$,
used while moving to the right, and allowing transitions between states of~$Q_L$ and of~$Q_R$ while moving to the left and
to the right, respectively, but still forbidding state changes on rewritten cells and by keeping all the other restrictions.
This model, called \emph{almost strongly limited automata}, still characterizes the context-free languages.
Furthermore, the two deterministic context-free languages mentioned in the previous paragraph
can be easily recognized by almost strongly limited automata having only deterministic transitions.

It would be interesting to know if almost strongly limited automata are able to accept all deterministic context-free languages
without taking nondeterministic decisions.

\section{Conclusion}

We discussed some restricted versions of one-tape Turing machines characterizing context-free languages.
Some other interesting models are presented in the literature. We briefly mention some of them.
 
In 1996 Jancar, Mr\'az, and Pl\'atek introduced \emph{forgetting automata}~\cite{JMP96}.
These devices can erase tape cells by rewriting their contents with a special
symbol. However, rewritten cells are kept on the tape and are still considered during the computation.
For instance, the state can be changed while visiting an erased cell. 
In a variant of forgetting automata that characterizes context-free languages,
when a cell which contains an input symbol is visited while moving to the left, its content is rewritten,
while no changes can be done while moving to the right.
This way of operating is very close to that of strongly limited automata. However, in strongly limited
automata the rewriting alphabet can contain more than one symbol. Furthermore, rewritten cells
are completely ignored (namely, the head direction and the state cannot be changed while visiting them)
except in the final scan of the tape from the right to the left end-marker.
So the two models are different. For example, to recognize the set of palindromes, a strongly limited
automaton needs a working alphabet of at least~$3$ symbols while, by definition,  to rewrite tape cells forgetting
automata use only one symbol~\cite{Pi15}.

If erased cells are removed from the tape of a forgetting automaton, we obtain another computational model
called \emph{deleting automata}. This model is less powerful. In fact it is not able to recognize all
context-free languages~\cite{JMP96}.

\smallskip

Wechsung proposed another complexity measure for one-tape Turing machines called~\emph{return complexity}~\cite{We75,WB79}.
This measure counts the maximum number of visits to a tape cell, starting from the first visit which modifies the cell content.
It should be clear that return complexity~$1$ characterizes regular languages (each cell, after the
first rewriting, will be never visited again, hence the rewriting is useless). Furthermore, for each~$d\geq 2$,
return complexity~$d$ characterizes context-free languages. Notice that this measure is dual with respect to the
one considered to define limited automata.\footnote{Indeed, the maximum number of visits to a cell up to
the last rewriting, namely the measure used to define limited automata, 
is sometimes called~\emph{dual return complexity}~\cite{WW86}.} Even with respect to return complexity, there exists a
hierarchy of deterministic languages (cf.\ Theorem~\ref{th:powerlimited}(iii) in the case of limited automata). 
However, this hierarchy is not comparable with
the class of deterministic context-free languages. For instance, it can be easily seen that the set of
palindromes, which is not a deterministic context-free language, can be recognized by a deterministic machine 
with return complexity~$2$. However, there are deterministic
context-free languages that cannot be recognized by any deterministic machine with return complexity~$d$, 
for any integer~$d$~\cite{Pe77}.

\smallskip

With the aim of investigating computations with very restricted resources, Hemaspaandra, Mukherji, and Tantau 
studied \emph{one-tape Turing machine with absolutely no space overhead}, a model which is very close to ``realistic'' computations, 
where the space is measured without any hidden constants~\cite{HMT05}.
These machines use the binary alphabet~$\Sigma=\{0,1\}$ (plus two end-marker symbols) and only the
portion of the tape which at the beginning of the computation contains the input.
Furthermore, no other symbols are available, namely only symbols from~$\Sigma$ can be used to rewrite the tape.
Despite these strong restrictions, there machines are able to recognize in polynomial time all context-free languages over~$\Sigma$. 

\subsection*{Acknowledgment}
Many thanks to Lane Hemaspaandra, editor of the SIGACT News Complexity Theory Column, for the careful reading of the
first draft of the paper. His suggestions and recommendations were very useful for improving the quality of the paper.

\bibliographystyle{splncs03}
\bibliography{sigact}

\end{document}